\documentclass[twoside,twocolumn]{article}
\usepackage[super,sort&compress,comma]{natbib} 
\usepackage{mhchem}
\usepackage{times,mathptm}
\usepackage{sectsty}
\usepackage{balance} 

\usepackage{graphicx} 
\usepackage{lastpage}
\usepackage[format=plain,justification=raggedright,singlelinecheck=false,font=small,labelfont=bf,labelsep=space]{caption} 
\usepackage{fancyhdr}
\begin{document}

\thispagestyle{plain}
\fancypagestyle{plain}{
\renewcommand{\headrulewidth}{1pt}}
\renewcommand{\thefootnote}{\fnsymbol{footnote}}
\renewcommand\footnoterule{\vspace*{1pt}%
\hrule width 3.4in height 0.4pt \vspace*{5pt}} 
\setcounter{secnumdepth}{5}

\makeatletter 
\def\subsubsection{\@startsection{subsubsection}{3}{10pt}{-1.25ex plus -1ex minus -.1ex}{0ex plus 0ex}{\normalsize\bf}} 
\def\paragraph{\@startsection{paragraph}{4}{10pt}{-1.25ex plus -1ex minus -.1ex}{0ex plus 0ex}{\normalsize\textit}} 
\renewcommand\@biblabel[1]{#1}            
\renewcommand\@makefntext[1]%
{\noindent\makebox[0pt][r]{\@thefnmark\,}#1}
\makeatother 
\renewcommand{\figurename}{\small{Fig.}~}
\sectionfont{\large}
\subsectionfont{\normalsize} 

\fancyfoot{}
\fancyfoot[RO]{\footnotesize{\sffamily{1--\pageref{LastPage} ~\textbar  \hspace{2pt}\thepage}}}
\fancyfoot[LE]{\footnotesize{\sffamily{\thepage~\textbar\hspace{3.45cm} 1--\pageref{LastPage}}}}
\fancyhead{}
\renewcommand{\headrulewidth}{1pt} 
\renewcommand{\footrulewidth}{1pt}
\setlength{\arrayrulewidth}{1pt}
\setlength{\columnsep}{6.5mm}

\twocolumn[
  \begin{@twocolumnfalse}
\noindent\LARGE{\textbf{Pair interactions between complex mesoscopic particles from Widom's particle-insertion method}}
\vspace{0.6cm}

\noindent\large{\textbf{Bianca M. Mladek$^{\ast}$\textit{$^{a}$}  and Daan Frenkel\textit{$^{a}$}}}\vspace{0.5cm}


\noindent \textbf{\small{DOI:10.1039/C0SM00815}}
\vspace{0.6cm}

\noindent \normalsize{We demonstrate that Widom's particle insertion
  technique provides a convenient and efficient method to determine
  the effective pair interaction between complex, composite
  soft-matter particles in the zero-density limit. By means of three
  different test systems, i.e.~amphiphilic dendrimers, electrostatic
  polymers and colloids coated with electrostatic polymers, we
  demonstrate the validity and the power of the presented method.

}
\vspace{0.5cm}
 \end{@twocolumnfalse}
  ]

\section{Introduction}
Soft materials, be they colloids, polymers or proteins, are often
complex constructs that live in a bath of small molecules.  In
addition, these mesoscopic particles themselves are often composite
objects that contain flexible moieties\footnote[2]{In what follows, we
  use the term ``mesoscopic'' to describe particles that contain
  internal degrees of freedom that will be integrated out in a more
  coarse-grained description.  Usually, these particles will be in the
  mesoscopic size range from nano- to micrometers.}.  As a result,
atomistic modeling of the structure and dynamics of soft materials
would need to span a wide range of length and time scales, which makes
such an approach infeasible in all but the simplest cases (few
mesoscopic particles and short times).

\footnotetext{\textit{$^{a}$~Department of Chemistry, University of
    Cambridge, Lensfield Road, Cambridge, CB2 1EW, United
    Kingdom. Tel: 0044 1223 33 63 53; E-mail: bmm32@cam.ac.uk}}

In an effort to simplify the description of the system at hand,
coarse-grained models are being developed that aim to capture the
mesoscopic and macroscopic behavior of the system by including in the
description only the most relevant degrees of freedom. In the simplest
coarse-graining approach each complex, mesoscopic particle is
characterized by only one effective coordinate. In the case of
particles that have on average inversion symmetry, such as
e.g. polymer-coated colloids, this is the coordinate of the center of
inversion symmetry.  For systems without such symmetry (or, to be more
precise, where this symmetry center cannot be identified with a fixed
coordinate in the molecular frame) it is conventional to use the
center of mass to specify the position of the particle, which is the
procedure that is followed for polymers or proteins.  The mesoscopic
particle is then represented as a ``soft'' sphere and the effective
interaction potential determines the softness. The simplest effective
potentials are obtained in the low-density limit where many-body
interactions can be ignored.  In spite of their simplicity, such
models have been shown to work successfully for a range of mesoscopic
systems (see e.g.\cite{Lik01,Bol01,Jus02,Den03,Goe04,Mla08,Cap10}).

Computing effective interactions for arbitrarily complex mesoscopic
particles is one of the key steps in the development of a
coarse-grained model.  There are many ways in which these interactions
can be obtained. The present paper describes an approach that we found
to be considerably more efficient than other approaches that we
explored.

The paper is organized as follows: in Sec.~\ref{standard-techniques}
we define the effective interaction and explain standard Monte Carlo
(MC) techniques to determine it, highlighting their strengths and
their limitations. In Sec.~\ref{formalism} we adapt Widom's particle
insertion method to the present problem and show that it is an
efficient way to determine effective interactions. In
Sec.~\ref{technique} we explain how to implement this formalism and
test the method for three different model systems in Sec.~\ref{tests},
comparing the results from the different techniques. In the concluding
section (Sec.~\ref{conclusion}) we point out possible limitations of
the method that we present.

\section{Computing effective interactions: the  standard approach}\label{standard-techniques}
Consider two mesoscopic particles confined in a volume $V$ and
positioned at a distance $R_{12}$ between them. At infinite dilution,
the effective pair interaction $\Phi_{\rm eff}(R_{12})$ between these
particles is related to their radial distribution function\cite{Han06}
$g(R_{12})$ via
\begin{equation}\label{gr}
g(R_{12}) = \exp\left[-\beta \Phi_{\rm eff}(R_{12})\right],
\end{equation}
where $\beta=1/kT$ is the reciprocal
temperature\cite{Kru89,Goe04}. The radial distribution function,
$g(R_{12})$, can easily be measured during the simulation and has to
be normalised to 1 for large separations between the
particles. Eqn.~\ref{gr} thus offers a straightforward way to measure
the effective interaction within simulations. However, the repulsion
between two mesoscopic particles usually increases as the particles
approach each other, possibly reaching several $kT$ and configurations
where particles are that close are rare. As a result, the relative
error in $g(R_{12})$ at short distances will therefore be large, which
results in a concomitantly large error in $\Phi_{\rm eff}(R_{12})$.
In order to obtain an accurate estimate of $\Phi_{\rm eff}(R_{12})$
without wasting time on irrelevant, though easily accessible
configurations, it is therefore necessary to use a simulation
technique that samples also the regime where $\Phi_{\rm
  eff}(R_{12})\gg kT$.

One way to overcome this sampling bottleneck is to sample all relevant
values of $R_{12}$ more or less uniformly.  Conceptually, the simplest
approach to achieve this is to introduce an external biasing potential
that acts on the effective coordinates and counteracts the repulsion
between the particles. The total interaction is then given by $\tilde
\Phi (R_{12}) = \Phi_{\rm eff}(R_{12}) + \Phi_{\rm bias}(R_{12})$ and
the radial distribution function is modified to read
\begin{eqnarray}
\tilde g(R_{12}) &\propto& \exp \left[ -\beta \tilde \Phi
  (R_{12})\right] \nonumber\\ &=& \exp \left[-\beta \Phi_{\rm
    eff}(R_{12})\right]\exp \left[-\beta \Phi_{\rm
    bias}(R_{12})\right].\nonumber
\end{eqnarray}
Comparing this last equation to Eqn.~\ref{gr}, we see that to
determine the radial distribution function of the {\it unbiased}
system, $g(R_{12})$, we have to increment the histogram of the
probability of finding the molecules separated by a certain distance
$R_{12}$ in each measurement by $\exp\left[\beta \Phi_{\rm
    bias}(R_{12})\right]$ rather than by 1. The effective interaction
can then be determined from this histogram via Eqn.~\ref{gr}.

However, there are two major disadvantages to this approach. First, to
sample all relevant distances uniformly, $\tilde \Phi (R_{12})$ should
be zero (or constant) for all $R_{12}$ and therefore the biasing
potential should ideally be $\Phi_{\rm bias}(R_{12}) = -\Phi_{\rm
  eff}(R_{12})$, which would correspond to knowing {\it a priori} the
sought-after answer.  In practice, long simulations are often needed
where the biasing potential is determined in a trial-and-error
procedure, enhancing the guess for $\Phi_{\rm bias}(R_{12})$ in a
rather cumbersome iterative process until the difference to
$-\Phi_{\rm eff}(R_{12})$ is smaller than $kT$, which will then allow
for an efficient sampling of all distances. But even once the biasing
potential is known with sufficient accuracy, sampling will still be
slow due to the fact that the mesoscopic particles have to diffuse
over the relevant range of $R_{12}$ values to visit each distance
often enough.

In an automated process, one can use the Wang-Landau
approach\cite{Wan01} to determine the biasing potential and thereby
the effective interaction in an iterative way. In this scheme,
$g(R_{12})$ is initially unknown and set to unity for all distances
$R_{12}$. During the simulation, moves from an old distance $R_{12}^o$
with potential energy $U(R_{12}^o)$ to a new distance $R_{12}^n$ with
potential energy $U(R_{12}^n)$ are accepted with a modified Metropolis
acceptance rule, where the probability to accept the move is given by
\[
P_{\rm acc}(R_{12}^o \rightarrow R_{12}^n) = \min\left[1,
  e^{-\beta\left[U(R_{12}^n)-U(R_{12}^o)\right]}\frac{g(R_{12}^o)}{g(R_{12}^n)}\right].
\]
In each measurement, the histogram for $g$ at the current position
$R_{12}$ is multiplied by a factor $f$, which is usually taken to be
2. At the same time, a histogram of the distances visited is measured
and once this histogram is sufficiently flat, a coarse guess for
$g(R_{12})$ has been obtained. To refine the results, $f$ is set to
$\sqrt{f}$, and the simulation is iterated like this until $f \sim
1$. Then, the effective interaction can be determined via
Eqn.~\ref{gr}.

Alternatively, one can break down the range of distances into several
small windows\cite{Tor77,Fre02} and then use simple biasing potentials
to force the system to stay within each window. Typical choices for
these so-called umbrella potentials are e.g.~hard walls\cite{Cha87} or
spring-like potentials $\Phi_{\rm bias}^j(R_{12}) = \frac{1}{2} k_j
(R_{12} - R_j)^2$, where the $k_j$ are spring constants that determine
the width of each window $j$ located at $R_j$. Carrying out separate
simulations for the different windows, one can systematically vary the
separations between the mesoscopic particles. Within each of these
windows $j$, a separate histogram of the probability of finding the
molecules a certain distance apart, $g^j(R)$, is recorded. At the end
of the simulations, the effective potential $\Phi_{\rm eff}(R_{12})$
between the molecules is obtained by merging the effective potentials
obtained within each window,
\[
\beta \Phi^j(R_{12}) = - \ln\left[g^j(R_{12})\right]-\beta \Phi_{\rm
  bias}^j(R_{12}) + c_j,
\]
where $c_j$ is a normalization constant. Since the various $c_j$ are
initially unknown, the concatenation of $\Phi_{\rm eff}(R_{12})$ will
display discontinuities at the windows' edges. To obtain a continuous
$\Phi_{\rm eff}(R_{12})$, the $c_j$ are chosen such that the data are
aligned to each other at the edges of the windows or, more
sophisticatedly, the multiple histogram method can be
implemented\cite{Fer88,Kum04,Fre02}.  Using this umbrella sampling,
care has to be taken to choose the windows and umbrella potentials in
a way that the variation of the effective interaction within each
window does not exceed 1-2 $kT$ to reliably sample the entire
window. Moreover, to improve quality of the matching of the different
parts of $\Phi_{\rm eff}(R_{12})$ at the edges of the windows, it is
advisable to use overlapping windows.

In view of the various limitations of above techniques, it would be
desirable to have a straightforward, unbiased method of sampling all
distances of approach between two mesoscopic particles with the same
probability, while obtaining an efficient estimate of the effective
interaction. We show in the next section how Widom's particle
insertion method can be used to achieve precisely this.

 
\section{Formalism}\label{formalism} 
The following formalism is based on Widom's insertion method to
calculate the chemical potential of non-uniform fluids\cite{Wid78}. We
adapt his approach to determine the effective interactions between two
mesoscopic particles in the zero density limit in a fast and
straightforward manner. The method presented here is also a
generalisation of the scheme used in Ref.~\citenum{Bol01}, where the
effective interaction between two self-avoiding, i.e.~hard-core,
polymers was determined from the overlap probability when placing the
polymers random distances apart. Despite this generalisation being
straight-forward, we are not aware of earlier implementations.

We consider a fluid of $N$ composite, mesoscopic particles confined in
a volume $V$. The effective coordinate of particle $i$ is denoted by
vector ${\bf R_i}$ with $i=1,\dots,N$.

The probability $P$ to find this system in a certain configuration
$\{{\bf R}_1, \dots, {\bf R}_N\}$ is given by the Boltzmann
distribution
\[
P\left(\{{\bf R}_1, \dots,{\bf R}_N\}\right) = \frac{1}{Z_N} e^{-\beta
  U\left({\bf R}_1, \dots,{\bf R}_N\right)}\; .
\]
$U$ is the potential energy of the system including all intra and
inter-particle interactions.  $Z_N$ is the configurational integral
given by
\[
Z_N = \int {\rm{d}}{\bf R}_1 \dots {\rm{d}}{\bf R}_N e^{-\beta
  U\left({\bf R}_1, \dots,{\bf R}_N\right)}.
\]
As Widom argued\cite{Wid78}, the density profile at position ${\bf r}$
within the volume can be written as
\[
\rho({\bf r}) = \frac{N}{Z_N} \int {\rm{d}}{\bf R}_1 \dots
    {\rm{d}}{\bf R}_{N-1} e^{-\beta U\left({\bf R}_1, \dots,{\bf
        R}_{N-1},{\bf r}\right)}.
\]
Splitting the potential energy in above equation into a contribution
due to all the interactions in the $(N-1)$-particle system and an
energy change, $\Delta U$, due to adding an $N$th particle at position
${\bf r}$, i.e.
\[
U\left({\bf R}_1, \dots,{\bf R}_{N-1},{\bf r}\right) = U\left({\bf
  R}_1, \dots,{\bf R}_{N-1}\right) + \Delta U\left({\bf R}_1,
\dots,{\bf R}_{N-1},{\bf r}\right),
\]
we can write that
\[
\rho({\bf r}) = \frac{N}{Z_N} \int {\rm{d}}{\bf R}_1 \dots
    {\rm{d}}{\bf R}_{N-1} e^{-\beta U\left({\bf R}_1, \dots,{\bf
        R}_{N-1}\right)} e^{-\beta \Delta U\left({\bf R}_1, \dots,{\bf
        R}_{N-1},{\bf r}\right)},
\]
which simplifies to
\[
\rho({\bf r}) = \frac{N Z_{N-1}}{Z_N} \left< e^{-\beta \Delta
  U\left({\bf R}_1, \dots,{\bf R}_{N-1},{\bf
    r}\right)}\right>_{N-1,{\bf r}},
\]
where $\left< x \right>_{N-1,{\bf r}}$ denotes an ensemble average
over quantity $x$ in the $(N-1)$ particle system at position ${\bf
  r}$.

Now, we apply this last formula to a system of only two mesoscopic
particles, labeled $1$ and $2$, whose effective coordinates are
positioned at ${\bf R}_1$ and ${\bf r} ={\bf R}_{2}$. Then,
\[
\rho({\bf R}_{2}) = \frac{N Z_{1}}{Z_2} \left< e^{-\beta \Delta
  U\left({\bf R}_1, {\bf R}_{2}\right)}\right>_{N=1,{\bf R}_{2}}.
\]
Since we want to determine a radially symmetric interaction potential,
we can position the first particle in the origin and place the second
at a distance $R_{12} = |{\bf R}_{2}-{\bf R}_{1}| = |{\bf
  R}_{2}|$. The above equation then simplifies to
\[
\rho(R_{12}) = \frac{N Z_{1}}{Z_2} \left< e^{-\beta \Delta
  U\left({R}_{12}\right)}\right>_{N=1,{R}_{12}}.
\]
The density profile can be expressed as $\rho({R}_{12}) = \rho
g({R}_{12})$, where $\rho$ is the average number density. Recalling
Eqn.~\ref{gr}, we find that
\begin{equation}\label{eff_int}
\beta \Phi_{\rm eff}({R}_{12}) = -\ln \left[\left< e^{-\beta \Delta
    U\left(R_{\rm 12}\right)}\right>_{N=1,{R}_{12}}\right] + c.
\end{equation}
 The constant $c$ cannot be directly measured since it is proportional
 to the configurational integral of the system. However, it is
 expected that the range of the effective interaction between two
 typical macromolecules is finite, i.e.~$\Phi_{\rm eff} \rightarrow 0$
 as $R\rightarrow \infty$. Therefore, with $R_{\rm max}$ being a
 separation between the mesoscopic particles where their interaction
 has decayed to zero, $c$ can be determined as
\[
c =\ln \left[\left< e^{-\beta \Delta U\left( R_{\rm
      max}\right)}\right>_{N=1, R_{\rm max}}\right].
\]

\section{Simulation Technique}\label{technique}
Eqn.~\ref{eff_int} allows us to determine the effective interaction
between two mesoscopic particles as a function of distance ${R}_{12}$
between their effective coordinates ${\bf R_{\rm i}}$, $i=1,2$, in a
straightforward, efficient and unbiased way.

Let the coordinates of the $M$ constituent parts of the mesoscopic
particles be given by $\left\{{\bf r}_{i,1},\dots,{\bf r}_{i,M}
\right\}$, $i=1,2$. In a first step, we equilibrate each of the two
mesoscopic particles in isolation and sample their configurational
space with a standard Metropolis MC simulation. The particle moves
employed will depend on the particular system under study and will be
chosen to ensure ergodicity. Each MC move from an old internal
configuration $o_i = \{{\bf r}_{i,1}^{o},\dots, {\bf r}_{i,M}^{o}\}$
to a new one $n_i=\{{\bf r}_{i,1}^n,\dots, {\bf r}_{i,M}^n\}$,
$i=1,2$, is accepted according to the Metropolis MC acceptance
rule\cite{Met53}, i.e.~with a probability
\[
P_{\rm acc}(o_i\rightarrow n_i) = \min\left(1,e^{-\beta
  \left[U(n_i)-U(o_i)\right]}\right),
\]
where $U(x_i)$ is the {\sl intra}molecular potential energy of
configuration $x$ of particle $i$.

Once the mesoscopic particles have been equilibrated, we randomly
sample from the equilibrated conformations that are used in the second
step of the algorithm. We fix the coarse-grained coordinate of the
first particle at the origin. We then generate effective coordinates
of the second particle, uniformly distributed in the interval
${R}_{\rm 12} \in [0,{R_{max}}]$ and we measure the {\sl
  inter}molecular potential energy $\Delta U\left(R_{\rm 12}\right)$
between the two mesoscopic particles\footnote[3]{To ensure good
  statistics at close approach of the mesoscopic particles, we sample
  all distances uniformly on a line segment. An alternative, but less
  accurate procedure would sample points uniformly distributed in a
  spherical volume, but such a procedure would disfavor the sampling
  of short distances.}. This procedure yields $e^{-\beta \Delta
  U\left(R_{\rm 12}\right)}$ as a function of $R_{\rm 12}$ with high
efficiency, since no acceptance test has to be applied in this last
step. Repeated sampling of steps one and two allows us to determine
the average in Eqn.~\ref{eff_int} and thereby the effective
interaction between the two molecules.


\section{Application to test systems}\label{tests}
To demonstrate the power and validity of the method presented above,
we determine the effective interactions of three different test
systems.

\subsection{Electrostatic polymers}\label{poly}
\begin{figure}[tb]
\centering
  \includegraphics[width=8cm]{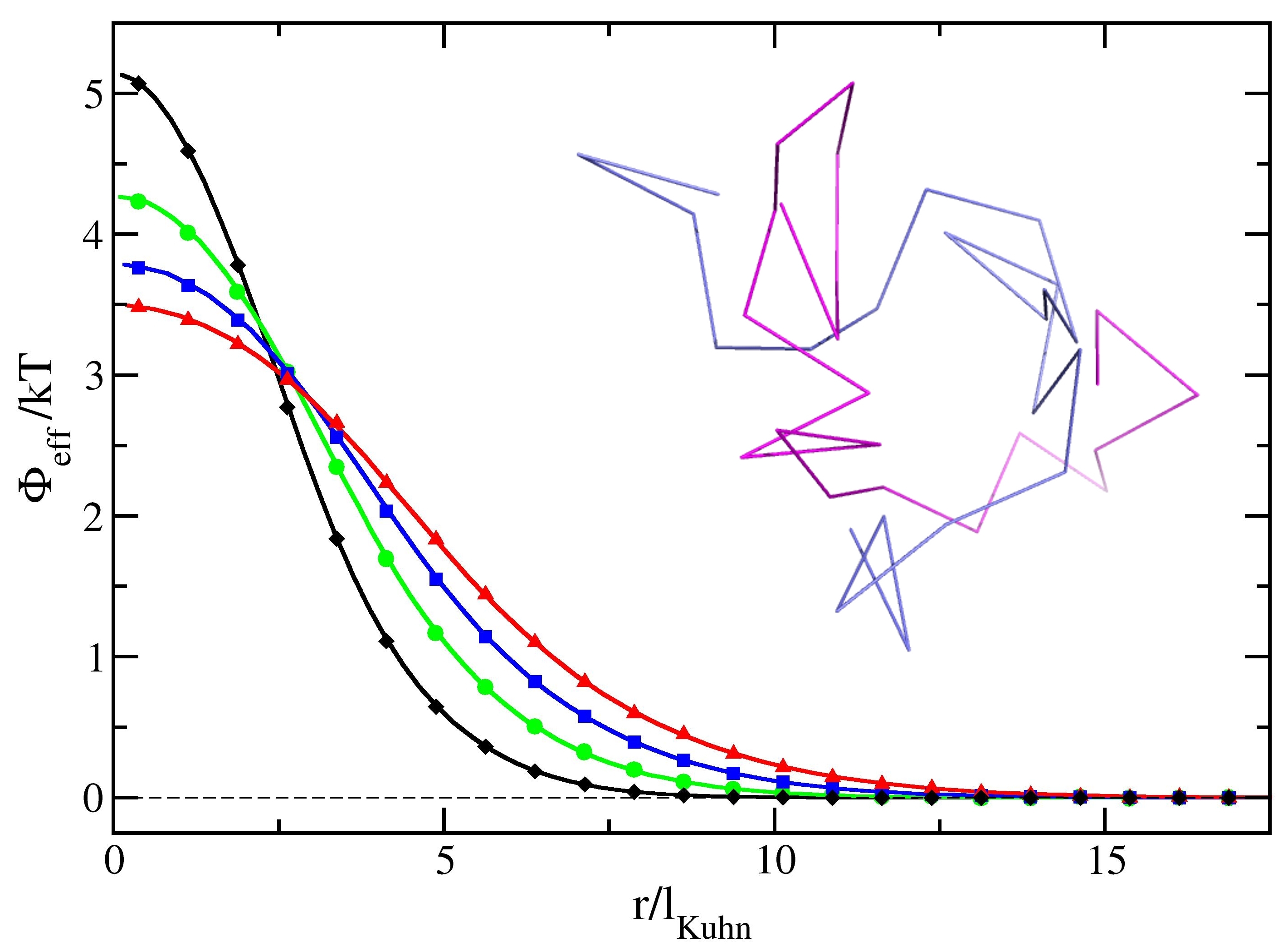}
  \caption{(Color online) Comparison between Widom's method (solid lines) and biased
    simulations (symbols) of the effective interaction $\Phi_{\rm eff}$
    between two electrostatically charged polymer chains of length $20$
    (diamonds), $30$ (circles), $40$ (squares) and $50
    l_{\rm Kuhn}$ (red, triangles) as a function of the distance between their
    centers of mass. The inset shows a simulation snapshot of two chains of
    $20 l_{\rm Kuhn}$.}
  \label{fgr:n20_polymer}
\end{figure}

As a first, simple test system, we study a strongly coarse-grained
representation of single-stranded DNA\cite{Zha01}, where the strands
are represented as electrostatically charged, freely jointed
chains. Vertices are separated by segments of fixed-length $l_{\rm
  Kuhn}=1.5$nm\cite{Smi96} and interact with each other via a
Debye-H{\"u}ckel potential which reflects both the charge of the
sugar-phosphate backbone of the DNA strands and the solvent
conditions\cite{Zha01},
\[
\beta \phi_{\rm DH}^{ij}  = q^2 l_B\frac{\exp(-\kappa r^{ij})}{r^{ij}},
\]
where $r^{ij}$ is the distance between vertices $i$ and $j$, $\kappa$
is the inverse Debye screening length, $q$ corresponds to the charge
per vertex and $l_B$ is the Bjerrum length. Here, we choose
$\kappa^{-1} = 0.67$nm, $l_B= 0.7$nm and $q = 3.05$e\cite{MlaXX} and
sample chains of 20, 30, 40 and 50$l_{\rm Kuhn}$ with both biased
simulations and Widom's particle insertion technique. In the former,
we sample the polymers for $1 \times 10^7$ MC sweeps, where each sweep
consists of 500 MC moves (i.e.~either a pivot move\cite{Lal69,Mad88},
crankshaft move\cite{Bau84}, rotation or translation of a
polymer). For Widom's method, we decorrelate each polymer for 500 MC
moves, using pivot and crankshaft moves, and with the two
configurations thus obtained we sample 500 different distances between
the polymers. We then repeat this cycle $2.5 \times 10^6$ times. In
Fig.~\ref{fgr:n20_polymer}, we compare the results from the two
methods and see that they perfectly coincide, confirming the validity
of Widom's method.

\begin{figure}[htb]
\centering
  \includegraphics[width=8cm]{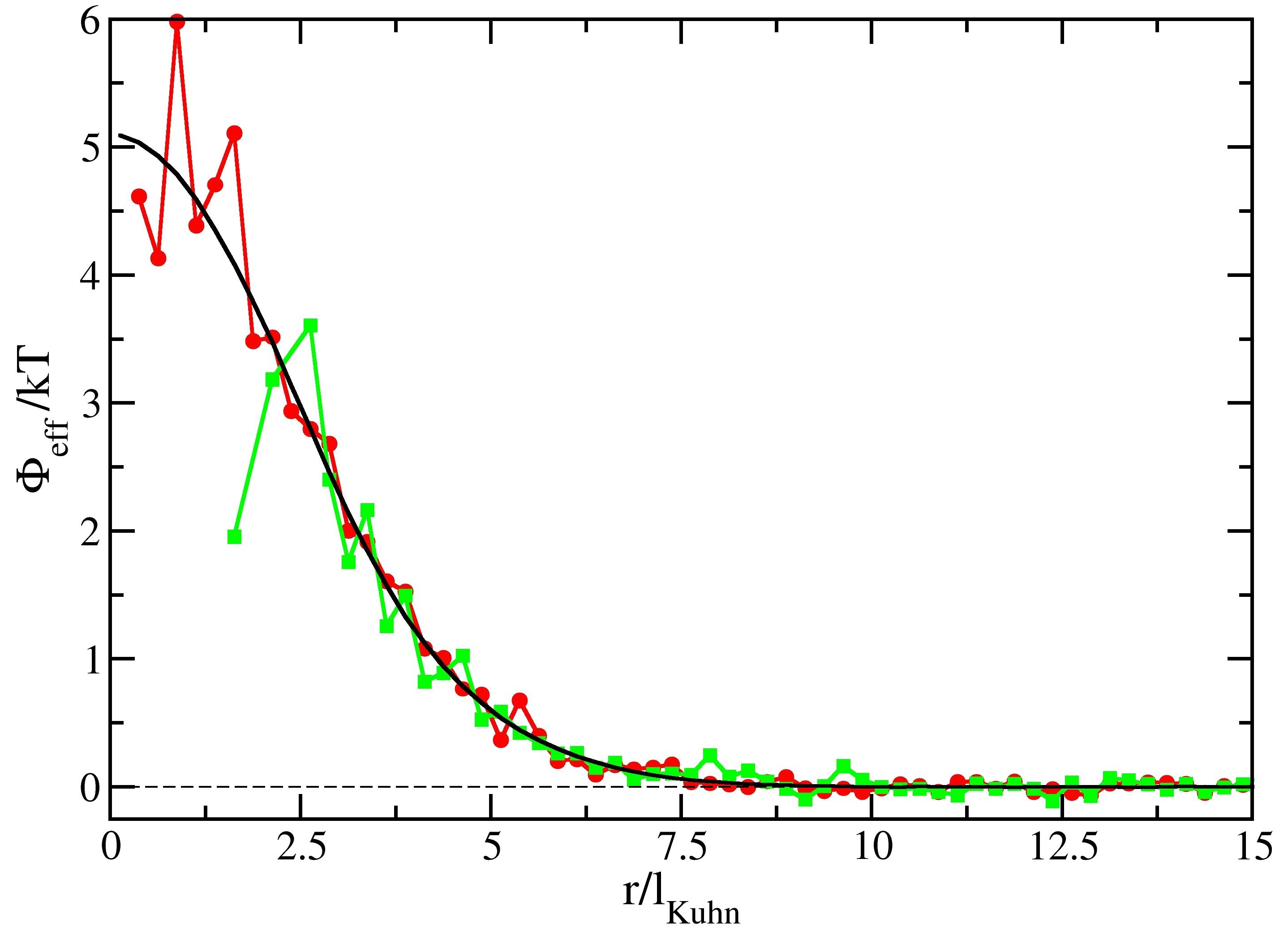}
  \caption{(Color online) Comparison between Widom's method (solid
    line), biased simulations (solid line with circles) and brute
    force calculations (solid line with squares) of the effective
    interaction $\Phi_{\rm eff}$ between two electrostatically charged
    polymers of length 20 after one hour of simulation time. While the
    data from Widom's insertion is equilibrated, statistics on the
    data of the other two methods is still insufficient.}
  \label{fgr:time_comp}
\end{figure}

To show how rapidly Widom's method converges, we compare to the
results from a brute force simulation and a biased one with $\Phi_{\rm
  bias} = -\Phi_{\rm eff}$. We ran all three simulations for the same
amount of computational time, i.e.~1 hour on a processor of our
quad-core dual Xeon (Harpertown) cluster. For the Widom method, this
corresponds to $6 \times 10^4$ cycles, for the brute force and the
biased simulation to $11 \times 10^4$ MC sweeps each. As shown in
Fig.~\ref{fgr:time_comp}, the results from Widom's insertion are
equilibrated already after this little of simulation time. By
contrast, the brute force simulation has not yet visited small
distances at all and the data of the biased simulation shows
considerable noise. This impressively highlights the efficiency of
Widom's method to determine effective interactions.

\subsection{Colloids coated with electrostatic polymers}
As a second example, we study a system of colloids coated with
electrostatically charged polymers that can move on the colloid's
surface. In this case, we determine the effective interaction between
the centers of the colloids rather than the centers of mass. The
vertices of the chains interact amongst each other as described in
Sec.~\ref{poly}, while the colloids are impenetrable spheres of fixed
radius $R_{\rm coll}$. Again, we implement crankshaft and pivot moves
for the polymer chains, but we also regrow them completely via a
configurational bias MC algorithm\cite{Fre02}, using 5 trial
directions in each step of the regrowth. The colloids are subject to
rotations and translations. Our model system is made of colloids of
$R_{\rm coll} = 4 l_{\rm Kuhn}$, each coated with 10 chains of 5 Kuhn
segments. For Widom's method, we simulated $4 \times 10^6$ cycles
consisting of 250 MC steps to decorrelate the coating of the colloids
before sampling 500 different distances between the colloids.

\begin{figure}[htb]
\centering
  \includegraphics[width=8cm]{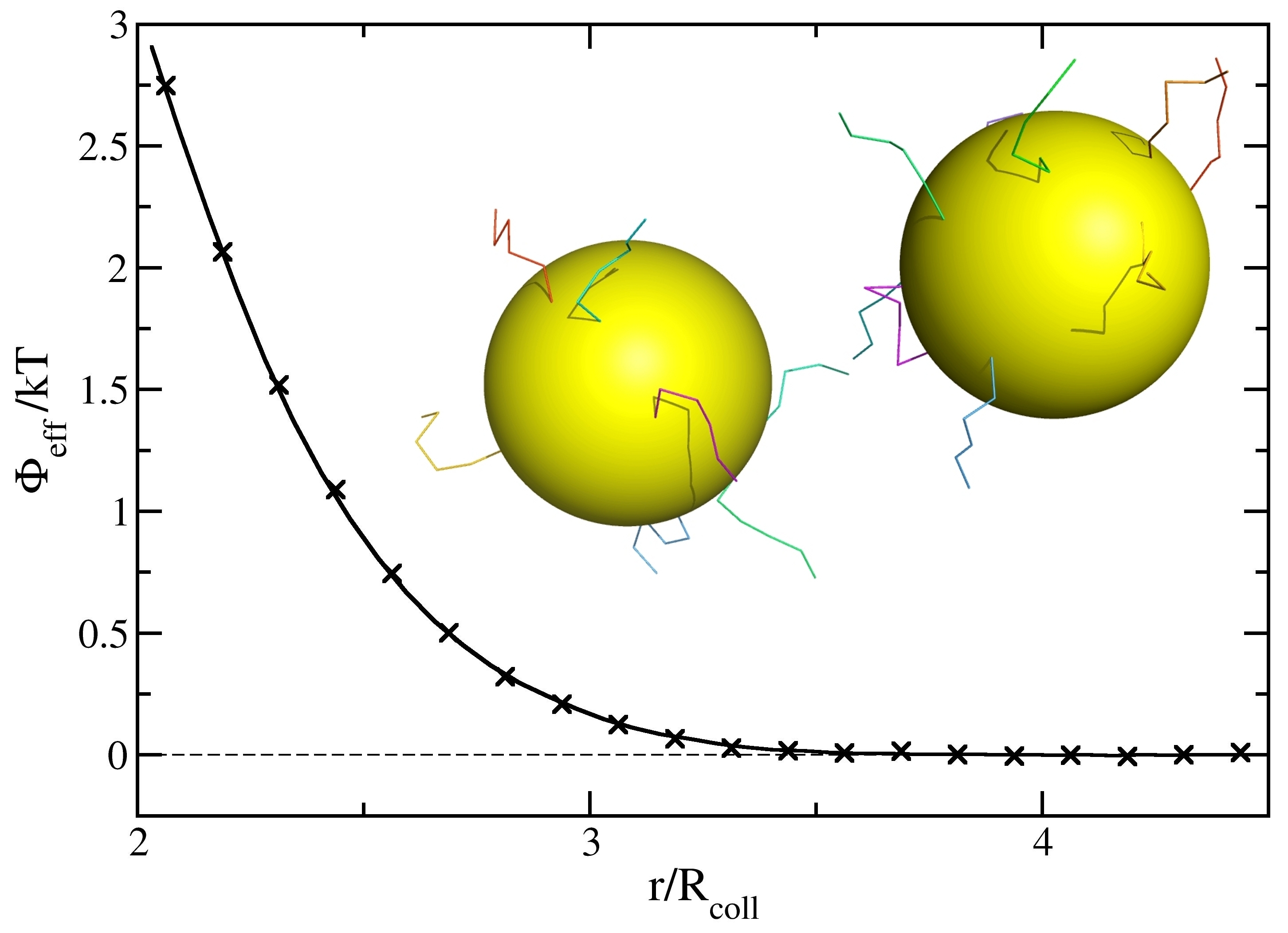}
  \caption{(Color online) Comparison between Widom's method (solid line) and brute force
    simulations (crosses) of the effective interaction $\Phi_{\rm eff}$
    between two colloids of radius $R_{\rm coll}= 4l_{\rm Kuhn}$  coated with 10
    electrostatically charged polymer chains of length $5 l_{\rm Kuhn}$ as a
    function of the distance between their centers. The inset shows a
    simulation snapshot of the colloids.}
  \label{fgr:moving_dna}
\end{figure}

In Fig.~\ref{fgr:moving_dna} we see that the Widom particle insertion
data perfectly reproduces the results of unbiased, brute force
simulations, which we ran for $1 \times 10^7$ MC sweeps. In the
latter, each sweep consisted of $220$ MC moves (i.e.~pivot or
crankshaft move, configurational bias MC regrowth, colloid rotation or
translation). While the brute force simulations needed two days of
simulation time and were averaged over 6 different simulation runs to
obtain reliable statistics, the data from a single simulation applying
Widom's method already shows decent statistics after as little as $5
\times 10^4$ cycles, which corresponds to 1.5 hours of simulation time
on a single processor of our cluster.

\subsection{Amphiphilic dendrimers}

Our final example concerns the determination of the effective
interaction between two amphiphilic dendrimers\cite{Mla08,Len09},
which are expected to show clustering behavior at sufficiently high
densities\cite{Mla08,Mla06,Mla08proc,LenXX}.

We sample two different second-generation amphiphilic dendrimers. They
have two central monomers and while the end-groups form a solvophilic
shell, all inner monomers represent the solvophobic core. The bonds
between monomers are modeled by the finitely extensible nonlinear
elastic potential, while all other interactions between monomers are
modeled by the Morse potential\cite{Wel98,Bal04,Mla08}. For the
parameters of the different interactions, we choose the same as for
dendrimer $D_5$ and $D_7$ from Ref.\citenum{Mla_phd} and we refer to this
reference for details.

Again, we implement Widom's insertion method and compare our results
to the effective interactions found in Ref.\citenum{Mla_phd} via umbrella
sampling. For the umbrella sampling, 15 slightly overlapping windows
with $R_{\rm max} = 5 R_g$ were used, where $R_g$ is the radius of
gyration of a single dendrimer. The systems were then sampled for $2
\times 10^8$ MC sweeps in each window. These simulations can be
carried out in parallel. For Widom's method, we sampled the dendrimers
for $4 \times 10^6$ MC cycles. In each of these cycles, each dendrimer
was decorrelated in 280 single monomer displacements and the resulting
configurations were used to sample 1000 different distances between
them.

\begin{figure}[htb]
\centering
  \includegraphics[width=8cm]{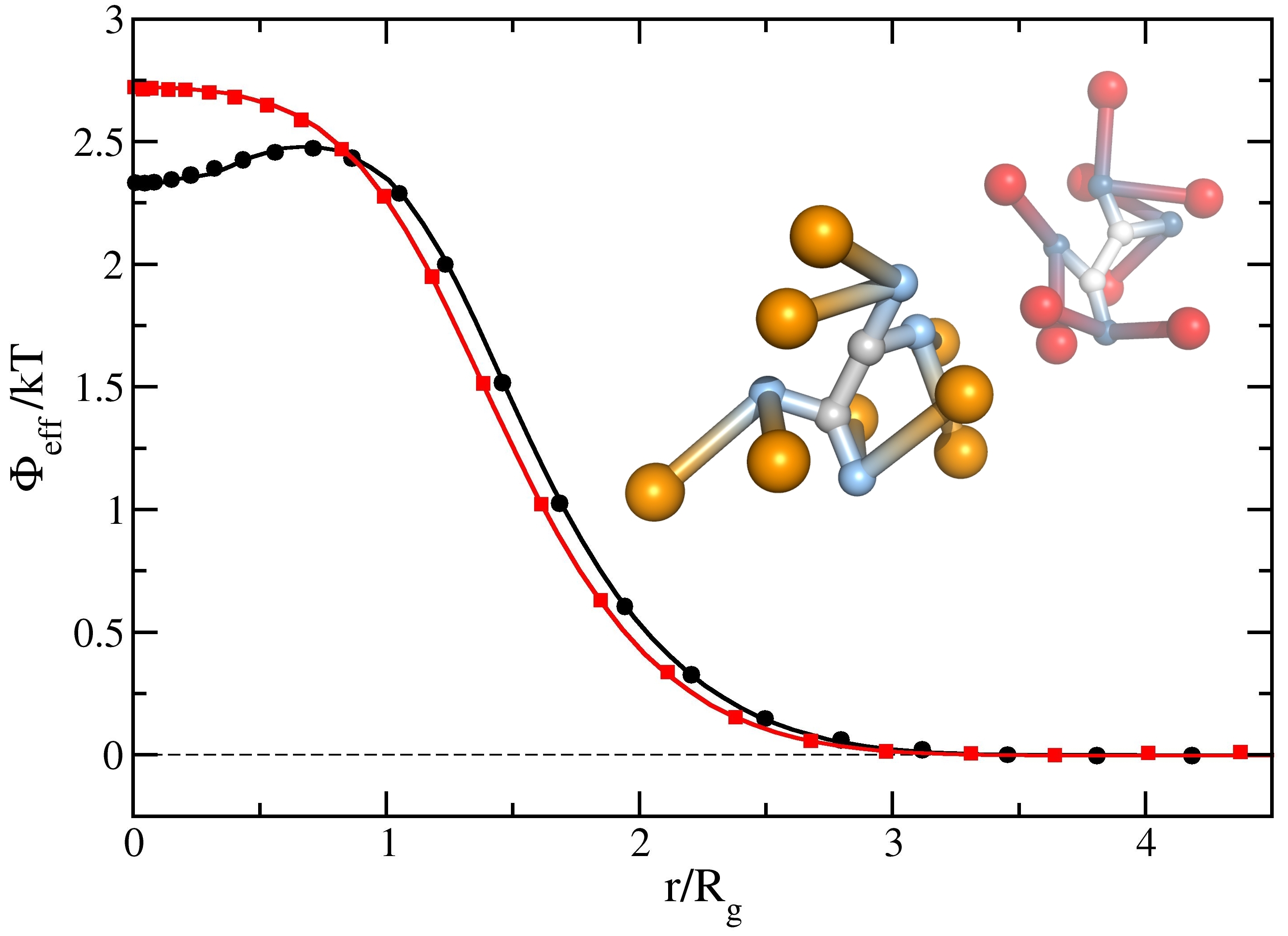}
  \caption{(Color online) Comparison between Widom's method (solid
    lines) and brute force simulations (symbols) of the effective
    interaction $\Phi_{\rm eff}$ between two amphiphilic $D_5$
    (circles) and $D_7$ (squares) dendrimers as a function of the
    distance between their centers of mass. The inset shows a
    simulation snapshot of two dendrimers.}
  \label{fgr:dendrimer}
\end{figure}

As can be seen from Fig.~\ref{fgr:dendrimer}, Widom's method manages
to capture the interaction potential even at close approach of the
dendrimers where deformations to the dendrimer's conformations are
expected. While Widom's technique already gives reliable statistics
for the effective interaction after as little as $5 \times 10^5$
cycles, corresponding to 3h of simulation time, umbrella sampling had
to be carried out on 15 processors for roughly a day to collect the
necessary statistics.

\section{Conclusions}\label{conclusion}
In this contribution we have shown that Widom's particle insertion
method can be adapted to determine the effective interaction between
two mesoscopic particles in a completely unbiased, efficient
way. While we showed the success of the method for three rather
different model systems, this method can not be applied in its present
form to particle insertion in an explicit solvent. Further, we expect
the method to fail for those systems where close approach between the
two particles leads to a very strong deformation of the molecules'
conformations that will not be sampled adequately in the ideal-gas
reference system that we use to generate independent
conformations. The usual way to test for the reliability or the
failing of Widom's method is the so-called overlapping distribution
method\cite{Shi82,Fre02}, where two simulations are needed: one of a
one-particle system, inserting a second one at frequent intervals; and
one of a two-particle system, removing one at frequent
intervals. Widom's method will only give reliable answers when the
probability distributions of finding certain energy differences
$\Delta U$ upon insertion and removal, respectively, have sufficient
overlap\cite{Fre02}. However, in these cases Widom's method will at
least allow for an educated guess of a biasing potential that can be
used with one of the standard techniques, and will thereby allow for a
considerable speed-up of the conventional methods.

In summary, the application of Widom's particle insertion method to
the determination of effective interactions between two mesoscopic
particles has several advantages over the standard techniques: first,
it allows for unbiased simulations, while the standard techniques
require a guess of appropriate biasing potentials to force the system
into close approaches. These potentials have to be determined with
high accuracy in an iterative process, or if less precise potentials
are used, simulations in multiple windows of different separation
ranges are required. Furthermore, Widom's method is very easy to
implement. Finally, biased or brute force simulations rely on the
diffusion of the particles through the simulation box, where each
change in separation will only be accepted with a certain probability
depending on the ratio between the Boltzmann weights of the old and
the new configurations. This makes the relative diffusion of highly
entangled particles slow and can lead to slow statistical fluctuations
in the data of the effective interaction. By contrast, within Widom's
method particles are simply placed at a given separation and the
intermolecular energy difference is measured, without having to employ
any acceptance/rejection step. Therefore, Widom's particle insertion
method immediately allows for an exploration of even the closest
approach and will give reliable statistics for the effective
interaction for all ranges of separation within very short simulation
times. For the cases that we have studied, this makes Widom's particle
insertion method the better technique to compute effective
interactions between mesoscopic particles.


\section{Acknowledgements} We thank S. Abeln from the Vrije Universiteit
Amsterdam and C.~N.~Likos from the University of Vienna for highly
stimulating discussions. BMM acknowledges funding from the EU via
FP7-PEOPLE-IEF-2008 No. 236663. DF acknowledges financial support from
the Royal Society of London (Wolfson Merit Award) and from the ERC
(Advanced Grant agreement 227758).

\balance
\footnotesize{
\bibliography{rsc} 
}

\end{document}